\documentstyle[aps,prl,array,multicol,epsf]{revtex}
\def\wide#1#2{
\postmulticols=0pt
\end{multicols}
\widetext
\noindent
\if#1t
\else
    \raisebox{9pt}[0in][0.0in]
    {$\rule{3.4in}{0.4pt}\rule{0.4pt}{6pt}$\hspace{3.6in}}
\fi
#2
\if#1b
\else
    \raisebox{-9pt}[0in][0.0in]
    {\hspace{3.55in}$\rule{0.4pt}{6pt}\rule[6pt]{3.5in}{0.4pt}$}
\fi
\setlength\premulticols{0pt}
\begin{multicols}{2}
\noindent
}

\begin{document}
\title{Hysteretic dynamics of domain walls at finite temperatures}
\author{T. Nattermann$^1$, V. Pokrovsky$^{2,3}$ and V.M. Vinokur$^4$}
\address{$^1$ Institut f\"{u}r Theoretische Physik, Universit\"{a}t zu 
K\"{o}ln,
D-50937 K\"{o}ln, Germany}
\address{$^{2}$Department of Physics, Texas A\&M University, 
College Station, 
TX 77843-4242, USA\\
$^{3}$Landau Institute for Theoretical Physics, Chernogolovka
Moscow
District, 142432, Russia}
\address{$^4$Argonne National Laboratory, 
Argonne, IL 60439, USA}
\date{\today}
\maketitle

\begin{abstract}
Theory of domain wall motion in a random medium is extended to the case when 
the driving field is below the zero-temperature depinning threshold and the 
creep of the domain wall is induced by thermal fluctuations. Subject to an ac 
drive, the domain wall starts to move when the driving force exceeds an
effective threshold which is temperature and frequency-dependent.
Similarly to the case of zero-temperature, the hysteresis loop displays
three dynamical phase transitions at increasing ac field amplitude $h_0$.
The phase diagram in the 3-d space of temperature, driving force amplitude
and frequency is investigated.
\end{abstract}


\begin{multicols}{2}

  Pinning dominated driven dynamics of elastic media in random
  environment is a paradigm for a vast diversity of physical systems.
  Examples include vortices in type II superconductors, charge density
  waves (CDW) in solids, stripe phases, Wigner crystals, dislocations
  in crystals, domain walls in magnets and many others \cite{DSF98}.
  Having appeared first in the context of dislocation dynamics
  \cite{iv-87}, the scaling theory of glassy dynamic state of random
  elastic media came to fruition in the context of CDW \cite{natt} and
  vortex lattices in high temperature superconductors
  \cite{colcreep,natt}, and enjoyed an impressive success in
  explaining a wealth of phenomenology of the low temperature vortex
  state\cite{blatt}. A closely related subject is the zero temperature
  depinning transition first studied for CDWs \cite{fisher,nar-fish92}
  and domain walls \cite{NST,nar-fish}.  Despite the significant
  recent progress, several key questions specific to glassy dynamics
  are yet poorly understood. One of such fundamental key issues,
  although known and extensively studied for more than hundred years
  in magnets, is hysteresis of interfaces subject to the applied ac
  drive and related aging and memory effects. A quest for urgent
  progress in understanding hysteretic behavior of magnetic domain
  walls is motivated also by emerging technological nano-scale
  magnetic systems whose ac properties are controlled by the
  hysteretic dynamics of interfaces.
\par
A step towards theoretical description of hysteretic behavior of
disordered interfaces has been undertaken in \cite{lnp-99}, where the
cyclic motion of the domain wall at zero temperature under the ac
field was investigated and the resulting magnetization hysteretic loop
was described.  
A finite temperature may change drastically the interface dynamics:
thermally activated creep motion becomes possible at any small drive.

In this Letter we develop a unified description of thermally activated
and over-threshold domain wall dynamics in {\it impure} magnets.  
We demonstrate that at
finite temperature new scales of length, activation energy and force
appear leading to emergence of a new, temperature and
frequency-dependent threshold field in the case of ac drive. The
latter is the first in a series of dynamical phase transitions.  To be
specific, we will speak on magnetic domain walls.  Accordingly we will
be using either of terms ''force'' or ''field'' equivalently.

{\it Finite temperature dc dynamics} 
The essential of the zero-temperature
dynamic behavior of an elastic medium in a random environment is the
existence of the finite threshold depinning force $h_{p}$, separating
immobile at $h<h_{p}$ and sliding at $h>h_{p}$ states of the system. Near
the threshold the sliding velocity $v$ shows a critical behavior 
\cite{fisher,nar-fish92,NST,nar-fish} $v\sim (h-h_{p})^{\beta }$. 
At finite temperatures and $h\ll h_{p}$
thermally activated drift motion controlled by the static rugged energy
landscape occurs. The latter is governed by the interface free energy

\begin{equation}
{\cal H}=\int d^{D}x\left\{ \frac{1}{2}\Gamma (\nabla Z)^{2}+V({\bf x}%
,Z)-h\,Z({\bf x})\right\}  \label{eq:H}
\end{equation}
where $\Gamma $ is the interface stiffness, $h$ is the external driving
force and $V({\bf x},Z)$ is the random impurity potential. $D$- dimensional
vector ${\bf x}$ is the coordinate along the interface, and $Z$ is the
coordinate of the transverse interface displacement. In the following we
assume that the disorder average of the random potential vanishes. There are 
two different types of impurities, random bond (RB) and random field
(RF) type in terms
of magnetic models.  The RB potential
obeys the Gaussian statistics with:
\begin{equation}
\overline{V_{RB}({\bf x},Z)\,V_{RB}({\bf 0},0)}
=v^{2}l^{D+1}\,\delta ({\bf x})\,\delta
(Z)  \label{eq:V_RB}
\end{equation}
where $v^{2}_{\phantom{1}}=v_{0}^{2}c$. $v_0,\;c$ and $l$ denote the
strength, the concentration and the correlation length of the
impurity potential. 
In the RF case $V_{RF}({\bf x},Z)=
\int\limits_0^Zh({\bf x},Z^{\prime})\,dZ^{\prime}$ where
the RF $h({\bf x},Z)$ has properties similar to $V_{RB}({\bf x},Z)$.

The static interface in a random environment becomes rough.
Its roughness obeys the scaling law \cite{hh}: 
\begin{equation}
w^{2}(L)=\overline{\left( Z({\bf x})-Z({\bf 0})\right) ^{2}}%
\approx l^{2}\left( \frac{L}{L_{p}}\right) ^{2\zeta };\,L=|{\bf x}
|  \label{eq:displacement-Z}
\end{equation}
where the roughness exponent is $\zeta =\frac{4-D}{3}$ for RF
and $\zeta \approx 0.2083(4-D)$ for  $4-D\ll 1$ and $\zeta =2/3$ for $D=2$, 
for RB impurities, respectively
\cite{fish86,hh}.
The rough
configuration develops over length scales $L\geq L_{p}$, where the Larkin
length $L_{p}$ is a distance at which a typical fluctuation of the pinning
forces, balanced by elastic forces, produces the transverse displacement $%
w\sim l$: 
\begin{equation}
L_{p}\approx l\left( {\Gamma }/{v}\right) ^{2/(4-D)
}\,.  \label{eq:L_cRF}
\end{equation}
Note that the typical slopes $w(L)/L$ of the wall 
vanish for $L\gg L_p$ since $\zeta < 1$. 
The energy barriers which
must be overcome to depin a segment of the wall with the linear size $L$ is 
\cite{iv-87}:
\begin{equation}
E_{B,0}(L)\approx T_p(L/L_{p})^{\chi};\,\,\,\chi = D-2+2\zeta .
\label{eq:E_B(L)}
\end{equation}

\noindent Here $T_{p}\equiv E_{B,0}(L_{p})\approx \Gamma l^{2}L_{p}^{D-2}$ is 
a typical pinning energy on a scale $L_{p}$. At temperature $T>T_{p}$ the 
effective force necessary for depinning drops rapidly with the 
temperature\cite{iv-87}. 
If an external driving force $h$ is applied, the total energy barrier 
$ E_B(L)$ becomes

\begin{eqnarray}
E_B(L) & \approx & E_{B,0}(L) - hL^{D}w(L)\nonumber\\
 & \approx &
 E_{B,0}(L) \left(1 - ({L}/{L_h})^{2-\zeta}\right) 
\label{eq:force}
\end{eqnarray}

\noindent where $L_{h}\approx L_{p}(h_{p}/h) ^{1/(2-\zeta )}$.
The energy barrier reaches its maximum $E_{B,max}$ at  
$L \approx (\frac{\chi}{D+\zeta})^{1/(2-\zeta)}L_h$ which gives 
$ E_{B,max}\approx T_p(h_p/h)^\mu$ and $\mu=\chi/(2-\zeta)$.

For completeness we note here that  equilibrium RF systems  in 
$d=D+1=2$ bulk dimensions have no  long range order at length scales 
$L\gg \xi_{RF}\approx l\exp{c(\Gamma/v)^{4/3}}$\cite{Villain}. For weak 
disorder $\xi_{RF}$ is  very large and can easily 
exceed the system size. 
As soon as an external field is applied such that $L_h <\xi_{RF}$, domain wall 
motion is dominated by the forces on scales $L_h$ and the absence of true 
long range order can be ignored.

As first found by Middleton \cite{middl}, close to $h_p$ the smallest energy 
barrier vanishes as $(h-h_p)^\theta$ with $\theta=2$. We therefore get  the 
effective energy barrier at $h\lesssim h_p$ if we 
replace in the above expressions $T_p$ by
$\tilde T_p = T_p(\frac{h_p-h}{h_p})^{\theta}$.

The time scale to overcome this barrier is of the order 
$\tau (L_{h})\approx \tau _{0}\,\exp \left( \tilde E_{B,max}/T\right) $, 
where $%
\tau _{0}$ is a microscopic hopping time, which leads to an average
velocity: 
\begin{equation}
v(h)\approx \gamma h\exp \left[ 
-\frac{ T_{p}}{T}\left(\frac{h_p-h}{h_p}\right)^{\theta}\left( \frac{h_{p}}{h}%
\right) ^{\mu }\right] \,
\label{eq:v(h)}
\end{equation}
where $\gamma $ is the effective friction coefficient 
\cite{iv-87,natt,radz+}.

At low temperature $T\ll T_{p}$ the dynamic threshold $h_{p}$ separates the creep
regime from the active sliding regime. As can be seen  from (\ref{eq:v(h)})
a characteristic
crossover field $h_T$ plays the role of the depinning force, where

\begin{equation}
\frac {h_{T}}{h_p}=\left(\frac{\tilde T_{p}(h_T)}{T}\right) ^{1/\mu}=
\left(\frac{T_{p}}{T}(1-\frac {h_T}{h_p})^{\theta}\right) ^{1/\mu }\,
\label{eq:h_T}
\end{equation}
\noindent At $h\approx h_{T}$, the
drift velocity increases rapidly and at larger fields it displays almost
linear behavior $v\approx \gamma h$. Note that $h_T$ is a monotonously 
decreasing function of temperature with  a maximum  $h_T=h_p$ at $T=0$. 

In a close vicinity of the threshold field $h_{p}$, the effective
energy barrier becomes small and even small thermal fluctuations may
be sufficient to overcome it. At finite temperatures and $h\lesssim
h_{p}$ the wall moves via thermal activation process with velocity
given by eq. (\ref{eq:v(h)}).  Strictly speaking, it means that at
finite temperatures the critical point shifts from $h=h_{p}$ to $h=0$.
Yet there remains a memory of the critical behavior around $h\approx
h_{p}$ displaying itself in a crossover behavior at finite but low
temperatures. The crossover is seen as a {\it rounding} of the $h-v$
characteristics
$v(h\approx h_{p})\sim T^{\beta /\theta }$
. We now can
write an interpolation formula for the velocity which is valid in a wide range
of variables:
\begin{equation}
v(h,T)=\gamma hF(x,y);\ x=h/h_{p};\ y=T_{p}/T,  \label{eq:v-inter}
\end{equation}
\begin{eqnarray}
\label{eq:F(x,y)}
F(x,y)&=&\frac{\Theta(1-x)}{ 1+(yx^{-\mu })^{\beta /\theta} }\exp \left[ -
yx^{-\mu }\left( 1-x\right)^{\theta}\right]\\
&&+\Theta (x-1)\left[\frac{1}{1+(yx^{-\mu})^{\beta/\theta}}+\left(1-\frac {1}{x}
\right)^{\beta}
\right]\nonumber
\end{eqnarray}
\noindent Here $\Theta (x)$ is the step function equal to zero at $x<0$ and
equal to 1 at positive $x$.
The interpolation formula (\ref{eq:F(x,y)}) satisfies
following requirements: (i) $v(h,T)=\gamma h$ at any fixed $T$ 
and $h\gg h_{T}$; 
(ii) $v(h,T)=\gamma h\exp \left[ -\frac{T_{p}}{T}\left( \frac{h_{p}-h}{h_{p}}
\right) ^{\theta}\right] $ for $h_{p}-h\ll h_{p}$
and $T\ll T_{p}$; (iii) $v(h,T)\approx
\gamma h
\exp \bigg[-\frac{T_p}{T}\bigg( \frac{h_p}{h}\bigg)^{\mu}\bigg]$ for
$T\ll T_p$, $h\ll h_p$ and $E_{B,max}/T\gg 1$ ; (iv) $v(h,T)\approx\gamma h_p
(h/h_p -1)^{\beta}$
for $(h/h_p-1)\ll 1$ and $T\ll T_{p}(h/h_p-1)^{\theta}$; (v) $v(h_p,T)\approx
\gamma h_p(T/T_p){\beta /\theta}$ for $T\ll T_p$.

So far we assumed that the propagating interface is self-affine. This 
is confirmed by numerical simulations in $D>1$ interface dimensions
for systems with weak disorder \cite{robbins}.
In $D=1$ dimensions the situation is less transparent: in simulations 
which use 
a bounded distribution of random fields
the interface appears to be self-affine \cite{nowak} or faceted
\cite{robbins2} depending on whether lattice effects are avoided  or 
admitted, respectively. We ignore here the possibility 
of faceted growth which occurs only in systems with narrow magnetic domain 
walls. For an unbounded distributions of random fields 
however a percolative self-similar domain wall propagation was observed 
\cite{Drossel}. In the following we will always assume, that the 
random fields distribution is bounded such that the domain walls remain 
well defined. This is also confirmed by our earlier simulation outside of 
the critical region \cite{lnp-99}.

{\it Alternating fields}. If the external drive is 
oscillating with 
frequency $\omega $, $h=h_{0}\sin {\omega t}$, the barriers for which $%
\omega \,\tau (L)>1$ cannot be overcome during one cycle of the ac
field.  From the condition $\omega \,\tau =1$ we find a 
{\it new frequency and temperature dependent}
magnetic field $h_{\omega }$ which obeys
\begin{equation}
\frac {h_{\omega}}{h_p}=
\left(\frac{T_{p}}{T\Lambda}(1-\frac {h_\omega}{h_p})^{\theta}\right) ^{1/\mu }\,
\label{eq:L_omega}
\end{equation}
where $\Lambda =\ln {1/(\omega \tau _{0})}$. $h_{\omega}$ plays the 
role of the 
dynamic threshold. At low fields $h_{0}<h_{\omega }$ there is no macroscopic 
motion of the wall, its segments oscillate between the metastable states with
close energies giving rise to dissipation\cite{iv-87}. Drift of the
wall starts at $h_{0}>h_{\omega }$. We assume $\omega \tau_{0}\ll 1,$ so
that 
$h_{\omega }<h_{T}$.
Various regimes of domain wall motion are summarized in Fig.1.
\begin{figure}
\begin{center}
\epsfysize=1.5truein
\epsfbox{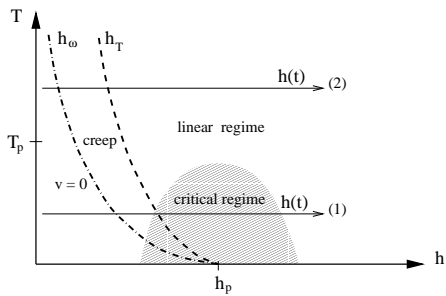}
\end{center}
\caption{The phase diagram for the domain wall hysteresis at finite
temperature and frequency. Solid and vertical lines  separate no sliding, 
thermal creep and mechanical drift regimes. See also explanations in the text.}
\end{figure}
Having derived the domain wall velocity as a function of the driving
field in the different $h-T$-regions we consider now the magnetic
hysteresis following from the motion of a single wall under the
influence of an oscillating field.  Since substantial length scales
are larger than $L_{p}$, where slopes are small, the domain wall will
be considered as a straight line (plane) characterized by one
coordinate $Z$ \cite {lnp-99}. Its dynamics is determined by equation
of motion

\begin{equation}
\dot{Z}=v(h(t))  \label{eq:main}
\end{equation}

\noindent 
 $Z$ varies between limiting values $0$ and $L$.
{\it Here $L$ is the linear size of the sample in the case of a
single domain wall or, in the  multi-domain case, equal to the average
 distance between expanding nuclei } . For harmonically oscillating field
$h=h_{0}$ $\sin \omega t$, equation (\ref{eq:main}) can be rewritten
in terms of $h$ only:

\begin{equation}
\frac{dZ}{dh}=\frac{v(h)}{\omega \sqrt{h_{0}^{2}-h^{2}}};\hspace{0in}\qquad
Z(h=0)=0  \label{eq:initial}
\end{equation}

\noindent Equations (\ref{eq:main}) and (\ref{eq:initial}) are valid for $%
h>h_{\omega }$. The field region $h<h_{\omega }$ where the motion has
zero drift velocity will not be considered here.  The value $h_{\omega
  }$ plays the same role as the threshold field $h_{p}$ plays at zero
temperature.  At $h>h_{\omega }$ the hysteresis is dominated by the
activation processes. In this respect it is similar to the nucleation
dominated hysteresis described in \cite{lnp-99} with two essential
differences. First, the activation relates to the formation of a
nucleus on the interface, not in the bulk. Second, the activation
energy depends on the magnetic field as a power function $\sim h^{-\mu
  }$ due to the distribution of barriers depending on their length
scale.

\begin{figure}
\begin{center}
\epsfysize=1.8truein
\epsfbox{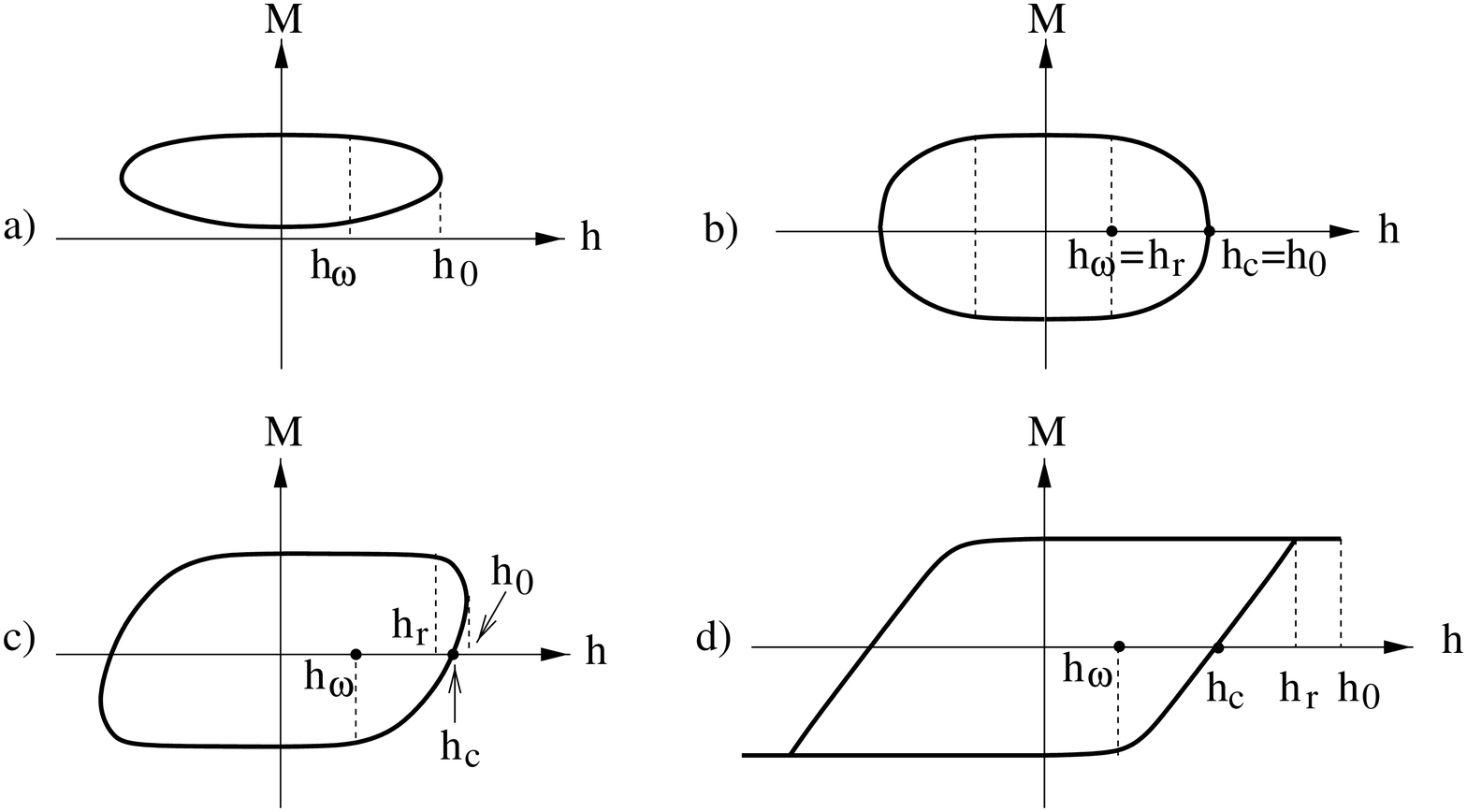}
\caption{Schematic pictures of hysteresis loops(HL). (a) Incomplete HL
for $h_0<h_{t1}$. (b) Symmetric HL for $h_0=h_{t1}$. (c) The HL for $h_t1<h_0<h_{t2}$. (d) The HL for $h_0>h_{t2}$. The values $h_p, h_c, h_r $ and $h_0$ 
are all marked in all figures}
\label{fig2}
\end{center}
\end{figure}

We start with large temperatures $T>T_{p}$ at which the
zero-temperature threshold field $h_{p}$ plays no role (field sweep
(2) in Fig.1). In this case the motion of the domain
wall is determined by equation (
\ref{eq:initial}) for $h>h_{\omega }$.
As in the case of zero temperature \cite{lnp-99} three dynamic phase 
transitions take place when the amplitude $h_0$ increases gradually at a fixed value of frequency. These transitions change the shape (symmetry) of the 
hysteresis loop. At the first of them proceeding at $h_0=h_{\omega}$ the 
hysteresis loop first appears, at smaller amplitudes $h_0 <h_{\omega}$ the 
magnetization remains unchanged. The hysteresis loop appearing at 
$h_0>h_{\omega}$ is characterized by incomplete magnetization reversal and 
reflection symmetry $h\rightarrow -h,\,\,\,M\rightarrow M$ as shown in 
Fig. 2a. This symmetry as well as incomplete magnetization reversal persists 
until the next dynamic phase transition at $h_0=h_{t1}$. 

At $h>h_{t1}$ the 
magnetization reversal becomes complete and hysteresis loop symmetry changes 
to inversion $h\rightarrow -h$, $M\rightarrow -M$ (see Figs. 2b, 2c). The 
value $h_{t1}$ is 
determined by a requirement that the domain wall proceeds from one sample 
boundary to another for half a period. At the next dynamic phase transition 
the symmetry of the hysteresis loop remains unchanged, but the part of the
cycle becomes reversible. Visually the hysteresis loop acquires 
characteristic "whiskers" as shown in Fig. 2d. The point of this transition
$h_{t2}$ is determined by a requirement that the domain wall proceeds from one 
sample boundary to another for quarter period.

Starting from the transition amplitude $h_0=h_{t1}$ each hysteresis loop goes 
through three important points. One of them is $h_{\omega}$, at which the 
motion of domain wall starts. Two others the so-called coercive field $h_c$ 
and reversal field $h_r$. At coercive field the magnetization turns into 
zero, at reversal field the magnetization becomes completely reversed (see 
Fig. 2b-2d). Note that for $h_0=h_{t1}$ $h_r=h_{\omega}$ and $h_c=h_0$; for
$h_0=h_{t2}$ $h_r=h_0$. All these fields can be found in our case. Equations 
for $h_{t1}$, $h_{t2}$ are:

\begin{equation}
\int_{h_{\omega }}^{h_{tn}}\frac{v(h)dh}{\sqrt{h_{tn}^{2}-h^{2}}}=\frac{%
n\omega L}{2}, n=1,2
\label{dynamic}
\end{equation}

For the case $T>T_{p}$ the equations (\ref{dynamic}) read:

\begin{equation}
g(x_n)=\frac{n\omega L}{2\gamma h_{T}}\,,\;
g(x)=\int\limits_{x_{\omega}}^{x}ye^{-y^{-\mu }}(x^{2}-y^{2})^{-1/2}dx
  \label{eq:ht}
\end{equation}

\noindent where $x_{n}=h_{tn}/h_{T}$, $n=1,2$; 
$x_{\omega }=h_{\omega
}/h_{T}=1/\Lambda ^{1/\mu }$. Thus, $x_{1}$, $x_{2}$ are functions of a
dimensionless parameter $u=\omega L/\gamma h_{T}$, where $L$ is the size of
the system or an average size of domains. Its asymptotic at small $u$
results in $x_{n}\approx [\ln (2/nu)]^{-1/\mu }$. 
The fields $h_{t1}$; $h_{t2}$ 
are close in this case: 
$(h_{t2}-h_{t1})/h_{t1}\approx \ln 2 /(\mu \ln u )$. The
requirement $h_{t1}>h_{\omega }$ is satisfied if $\gamma h_{T}\tau _{0}<L$.
The coercive field $h_{c}$ and the reversal field $h_{r}$ are determined by
equations: $M(h_{c})=0$; $\mid M(h_{r})\mid =M_{s}$.

A simple analysis at small $u$ results in $h_{c}\approx h_{t1}%
\text{; }h_{r}\approx h_{t2}$. The area ${\cal A}$ 
of the hysteresis loop at $u<<1$ and 
$h_{0}>h_{t1}$ does not depend strongly on the amplitude $h_{0}$ and 
becomes ${\cal A}\approx 4h_{r}{\cal M}_{s}$. Fig.2 
shows typical hysteresis loops and illustrates the geometrical 
meaning of the field $h_\omega, h_c, h_r$.  The dependence of
magnetization on magnetic field is given by equation: $M(h)=M_{s}\left( 
\frac{2Z(h)}{L}-1\right) $.

Finally, in the range of moderately low temperature 
$T<T_{p\text{ }}$ the
more complete expression (10) and (11) have  to be used in integrating 
equation (14). 

It interesting to note that the similar (dynamic) transition from
  incomplete to complete hysteresis 
was observed even in a standard simplistic mean-field model for pure magnets with the reaction described by the Brilloin function \cite{Chakrabarti}.  This suggests that this kind of dynamic transition we discuss, may be a generic property of nonlinear systems. Another note is in order: hysteresis in large multidomain magnet samples is a very complex phenomenon and cannot be always reduced to motion of a single domain wall (see for example numerical simulations of random Ising model 
in \cite{sethna}, where hysteretic and memory efects unlikely reducible to motion of a sible DW were revealed). 
 
In conclusion, we have investigated critical creep motion at low, $T\ll
T_{p}$, and at high, $T>T_{p}$, temperatures, $T_{p}$ being the depinning temperature, 
 and constructed the dynamic phase diagram. At low temperatures
creep at $h\approx h_{p}$ retains features of the critical behavior and
exhibits the rounding of the $h-v$ characteristic, according to \cite{middl}. 
At  finite frequencies $\omega $, a
new characteristic field $h_{\omega }<h_{p}$ comes into play, and the
transition from the sliding regime to pinning dominated activation motion is
shifted to $h_{\omega }$.

{\it Acknowledgements.} We are grateful to A. Middleton for a valuable
discussion.  One of us (V.P.) acknowledges the support of this work by
the NSF under the grant DMR-97-05182, DMR-00-72115 and by the DOE
under the grant DE-FG03-96ER45598 as well as from Humboldt-foundation.
The work at ANL was supported by the
by the  U.S. DOE, Office of Science under contract No. W-31-109-ENG-38.

\end{multicols}

\end{document}